\documentclass[aps,prl,twocolumn,groupedaddress,showpacs,showkeys]{revtex4}
\usepackage{natbib}
\usepackage[latin1]{inputenc}
\usepackage[dvips]{graphicx}
\usepackage{amsmath,amsbsy,amsfonts,amssymb}
\usepackage{bm}

\newcommand{\ket}[1]{\ensuremath{\left|#1\right>}}
\newcommand{\cafp}{\ensuremath{^{40}Ca^{+}}}
\newcommand{\srp}{\ensuremath{^{88}Sr^{+}}}
\newcommand{\bap}{\ensuremath{^{138}Ba^{+}}}
\newcommand{\bep}{\ensuremath{^{9}Be^{+}}}
\newcommand{\mgp}{\ensuremath{^{25}Mg^{+}}}
\newcommand{\omz}{\ensuremath{\omega_{z}}}
\newcommand{\state}[3]{\ensuremath{\,^{#1}{#2}_{#3}}}
\newcommand{\down}{\ensuremath{\ket{\downarrow}}}
\newcommand{\up}{\ensuremath{\ket{\uparrow}}}
\newcommand{\downplus}[1]{\ensuremath{\ket{\downarrow,{#1}}}}
\newcommand{\upplus}[1]{\ensuremath{\ket{\uparrow,{#1}}}}
\newcommand{\unit}[1]{\ensuremath{\,\rm #1}}
\begin{document}
\title{Trapped-Ion Quantum Logic Utilizing Position-Dependent ac Stark Shifts}
\author{Peter Staanum~\footnote{E-mail: staanum@phys.au.dk}}
\author{Michael Drewsen}
\affiliation{Department of Physics and Astronomy, University of
Aarhus, DK-8000 Aarhus C, Denmark.}
\date{\today}
\begin{abstract}
We present a scheme utilizing position-dependent ac Stark shifts
for doing quantum logic with trapped ions. By a proper choice of
direction, position and size, as well as power and frequency of a
far-off-resonant Gaussian laser beam, specific ac Stark shifts can
be assigned to the individual ions, making them distinguishable in
frequency-space. In contrast to previous all-optical based quantum
gates with trapped ions, the present scheme enables individual
addressing of single ions and selective addressing of any pair of
ions for two-ion quantum gates, without using tightly focused
laser beams. Furthermore, the decoherence rate due to off-resonant
excitations can be made negligible as compared with other sources
of decoherence.

\end{abstract}
 \maketitle
In recent years, physical realizations of quantum computers have
received growing interest. Several very different physical
implementations have been considered~\citep{Nielsen-Chuang}, and
schemes based on a string of trapped ions, as first introduced by
Cirac and Zoller~\citep{Cirac-Zoller}, are among the most
promising and popular candidates for demonstrating large scale
quantum logic. In these schemes two internal levels of an ion
represent a quantum bit (qubit), which can be manipulated through
laser interactions. Two key requirements are individual addressing
of the ions for single-qubit manipulations, and the ability to
make gate-operations between any pair of ions. Although multi-ion
entanglement was demonstrated in a recent
experiment~\citep{Sackett-entangled}, individual and selective
addressing remains a major experimental challenge for making an
ion-trap quantum computer. The difficulties of addressing
originates from the need for high trap-frequencies, to ensure
efficient motional ground-state cooling and high gate-speeds,
which leads to a small spatial separation of the ions. In
addition, the ion separation decreases with an increasing number
of ions~\citep{James}. In current experimental setups where
ground-state cooling has been demonstrated, typical
trap-frequencies are $\omz=2\pi\times 0.7\unit{MHz}$ in
Innsbruck~\citep{Leibfried-ICAP} and $\omz=2\pi\times
10\unit{MHz}$ at NIST~\citep{Turchette-heating}, which yields
minimum spacings of $7.1\unit{\mu m}$ (two \cafp-ions) and
$3.2\unit{\mu m}$ (two \bep-ions), respectively.

The most obvious method for individual addressing is simply to
focus a laser beam onto a single ion. This was demonstrated by the
Innsbruck-group~\citep{Naegerl-individual}, but it is extremely
demanding with the much stronger traps used at NIST or when more
ions are involved. A few more complex methods for individual
addressing of ions have been presented, which use either
position-dependent micromotion~\citep{Leibfried-individual,
Turchette-entangled} or a position-dependent magnetic
field~\citep{Mintert}, but they are either technically demanding
or hard to generalize beyond two ions.

\begin{figure}
  \centering
  \includegraphics[width=\linewidth]{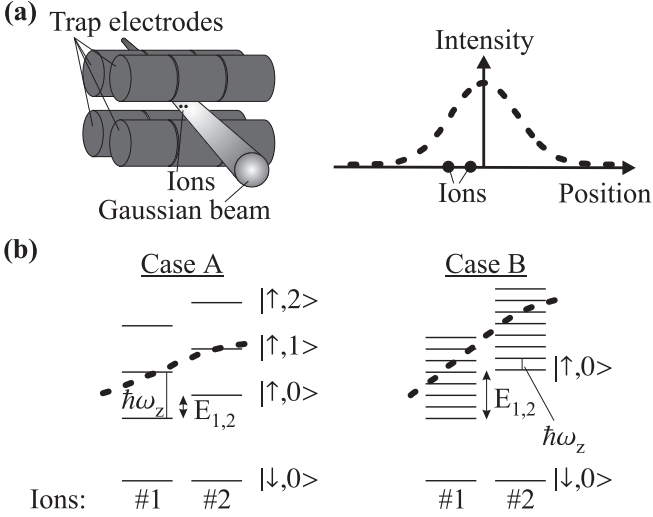}
  \caption{The basic idea of individual addressing. (a) Left: Sketch of
  the ac Stark-shifting laser beam and two ions in a linear Paul trap.
  Right: Position of the ions with respect to the intensity
  distribution of the laser beam. (b) The associated ac Stark-shifted
  energy levels of the two ions (assuming two internal states, $\down$ and $\up$)
  in the harmonic trapping potential of oscillation frequency \omz\
  (external states \ket{0}, \ket{1}, etc..). Individual addressing is
  considered in two cases. Case A: $\hbar\omz>E_{1,2}\gg\hbar\gamma_{res}$
  and Case B: $E_{1,2}\gg\hbar\omz\gg\hbar\gamma_{res}$. Note: For
  convenience, the energy-levels are shifted, such that the \down\
  state has the same energy for both ions.}
  \label{fig:idea}
\end{figure}

In this Letter, we propose to use a position-dependent
energy-shift, an ac Stark shift, of the qubit-levels, to obtain a
unique resonance-frequency of each ion, such that the ions can be
addressed individually just by tuning the frequency of a laser
beam illuminating the whole ion string. In addition, we
demonstrate how a position-dependent ac Stark shift can be used
for selecting any pair of ions in a multi-ion string for
implementing a two-ion quantum gate, e.g., a M\o lmer-S\o rensen
gate~\citep{Anders, Klaus-entanglement}. Our scheme is technically
not very demanding, since it relies on applying a far-off-resonant
ac Stark-shifting laser beam focussed to a spot size larger than
the ion spacing (see Fig.~\ref{fig:idea}). This feature makes the
scheme applicable even in experiments with relatively tightly
confining traps ($\omz/2\pi\sim 10\unit{MHz}$).

First, we consider the criteria for performing single qubit
operations between states of the type $\downplus{n}$ and
$\upplus{n'}$, where $\down$ and $\up$ are the two eigenstates of
the ions and $n$ is the vibrational quantum number for one of the
motional modes of the ions. To selectively manipulate such two
states of a single ion in a string, the spectral resolution
$\gamma_{res}$ of the laser performing the qubit operation has to
much better than the trap-frequency \omz\, and sufficiently high
that transitions in any other ion are prohibited. For simplicity,
in the following, we consider a two-ion string, with one motional
mode having the oscillation frequency \omz, and with the ac Stark
shift induced energy-difference between the two ions being
$E_{1,2}$ (see Fig.~\ref{fig:idea}). First, we treat the situation
where $\hbar\omz>E_{1,2}\gg\hbar\gamma_{res}$ as sketched in Fig.
1b (Case A). In this case, $E_{1,2}=\hbar\omz/2$ is the optimum
choice, since a laser resonant with a specific transition
$\downplus{n}\leftrightarrow\upplus{n'}$ in one ion, is maximally
off-resonant with all the transitions of the type
$\downplus{n}\leftrightarrow\upplus{n'}$,
$\downplus{n}\leftrightarrow\upplus{n'+1}$, or
$\downplus{n}\leftrightarrow\upplus{n'-1}$ in the other ion,
leading to the highest possible gate-speed. In the case
$E_{1,2}\gg\hbar\omz \gg\hbar\gamma_{res}$ (Case B in
Fig.~\ref{fig:idea}b), a laser resonant with a transition
$\downplus{n}\leftrightarrow\upplus{n'}$ in one ion, is only
resonant (or near-resonant) with a transition
$\downplus{n}\leftrightarrow\upplus{n'+m}$ in the other ion, where
$|m|\gg 1$. In the so-called Lamb-Dicke limit such a transition is
strongly suppressed~\citep{James, Stenholm}. Case B is
particularly interesting when more than two ions are present,
since even in such cases the gate-time will only be limited by the
vibrational frequency \omz\ instead of a fraction thereof as in
Case A.

\begin{figure}
  \centering
  \includegraphics[width=\linewidth]{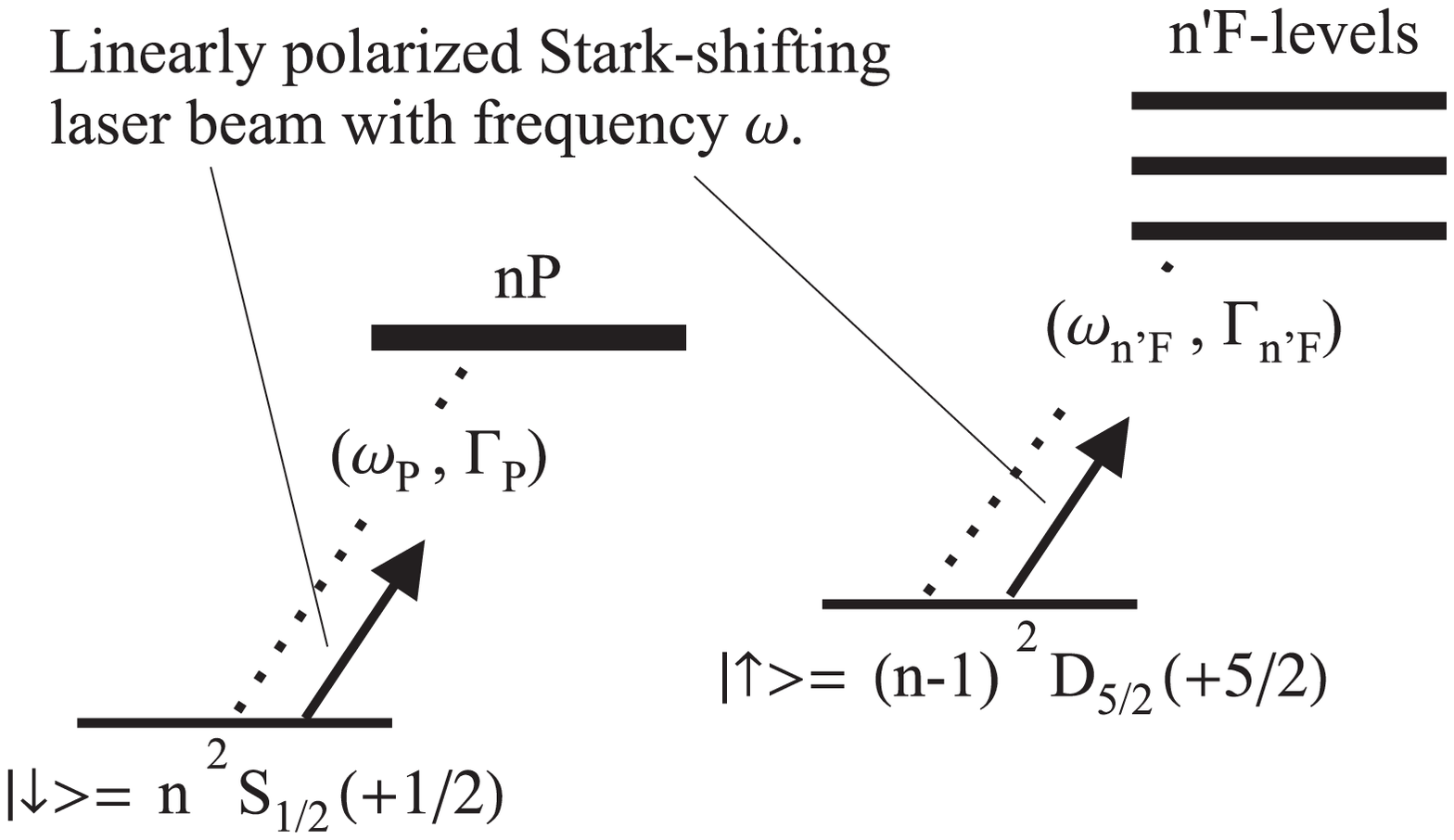}
  \caption{Relevant energy-levels and transitions in alkaline earth ions
  (e.g.,\cafp, \srp, and \bap) for calculating the ac Stark-shifts of the
  qubit states ($\down=\state{2}{S}{1/2}(m_{J}=+1/2)$ and
  $\up=\state{2}{D}{5/2}(m_{J}=+5/2)$) in the case of a linearly polarized,
  far-off-resonant laser beam. The Stark-shifting laser beam is assumed to be
  so far red detuned that the fine-structure splitting of the P- and F-levels
  can be neglected.}\label{fig:levels}
\end{figure}

An experimental realization of the above situation can be
achieved, e.g., by a string of two \cafp-, \srp-, or \bap-ions,
with the qubit states \down\ and \up\ represented by the
$\state{2}{S}{1/2}(m_{J}=+1/2)$ ground state and the
$\state{2}{D}{5/2}(m_{J}=+5/2)$ metastable state,
respectively~\footnote{Indeed the \state{2}{S}{1/2}- and
\state{2}{D}{5/2}-states in \cafp\ are used as qubit-levels in
Innsbruck~\citep{Leibfried-ICAP}.}. The far-off-resonant
Stark-shifting laser beam is set to propagate perpendicular to the
ion string, as indicated in Fig.~\ref{fig:idea}a, and its
polarization is assumed to be linear along the inter-ion axis.
Assuming a Gaussian intensity profile with a waist $W$, a maximum
difference in the ac Stark shift of the ions is obtained by
displacing the laser beam by $W/2$ with respect to the center of
the ion string. The relevant internal levels of the considered
ions, with respect to the Stark-shifting laser beam, are shown in
Fig.~\ref{fig:levels}. For simplicity, we assume that the
Stark-shifting laser beam is so far red detuned from any
transition-frequency that fine-structure splitting can be
neglected.

The ac Stark shift
$\varepsilon_{\uparrow}-\varepsilon_{\downarrow}$ of the
$\up-\down$ transition of a single ion can be calculated by
summing the contributions from all relevant dipole-allowed
couplings. The dominant shift of \down\ is from the $nS-nP$
coupling, whereas the shift of \up\ is composed of contributions
from a series of $(n-1)D-n'F$ couplings. This gives rise to the
following approximate expression for the ac Stark shift:
\begin{eqnarray}\label{eq:E-shift}
\varepsilon_{\uparrow}-\varepsilon_{\downarrow}
=&&\frac{3\pi c^{2}}{2}\Biggl[\frac{1}{\omega_{P }^{3}}
\left(\frac{\Gamma_{P}}{\omega_{P}-\omega}+\frac{\Gamma_{P}}
{\omega_{P}+\omega}\right)\nonumber\\
&&-\sum_{n'}\frac{1}{\omega_{n'F}^{3}}\left(\frac{\Gamma_{n'F}}
{\omega_{n'F}-\omega}+\frac{\Gamma_{n'F}}{\omega_{n'F}+\omega}
\right)\Biggr]I_{ion}\nonumber\\
&&\equiv\psi\times I_{ion},
\end{eqnarray}
where $\omega$ is the laser-frequency, $\omega_{P}$ and
$\omega_{n'F}$ are the $nS-nP$ and $(n-1)D-n'F$
transition-frequencies, $\Gamma_{P}$ and $\Gamma_{n'F}$ are the
corresponding spontaneous decay-rates, and $I_{ion}$ is the
intensity of the Stark-shifting laser beam at the position of the
ion~\citep{Grimm-opt}. $\psi$ is a parameter that depends only on
the properties of the ion and the laser-frequency. With the two
ions positioned at $r_{\pm}\equiv W/2\pm\Delta z/2$, where $\Delta
z=(e^{2}/2\pi\epsilon_{0}m\omz^{2})^{1/3}$ is the equilibrium
spacing of the ions of mass $m$, and given the laser
intensity-profile $I(r)=I_{0}\exp[-2r^{2}/W^{2}]$,
Eq.~\eqref{eq:E-shift} leads to the following difference in the
transition-frequency of the ions:
\begin{align}\label{eq:E-total}
&E_{1,2}=\kappa\left(\varepsilon_{\uparrow}-\varepsilon_{\downarrow}\right)
=\kappa\times\psi\times I_{0},
 \intertext{where}
 \label{eq:kappa}
&\kappa=2\sinh(\Delta z/W)\exp\left\{-\frac{1}{2}\left[1+(\Delta
z/W)^{2}\right]\right\}.
\end{align}

\begin{figure}
  \centering
  \includegraphics[width=\linewidth]{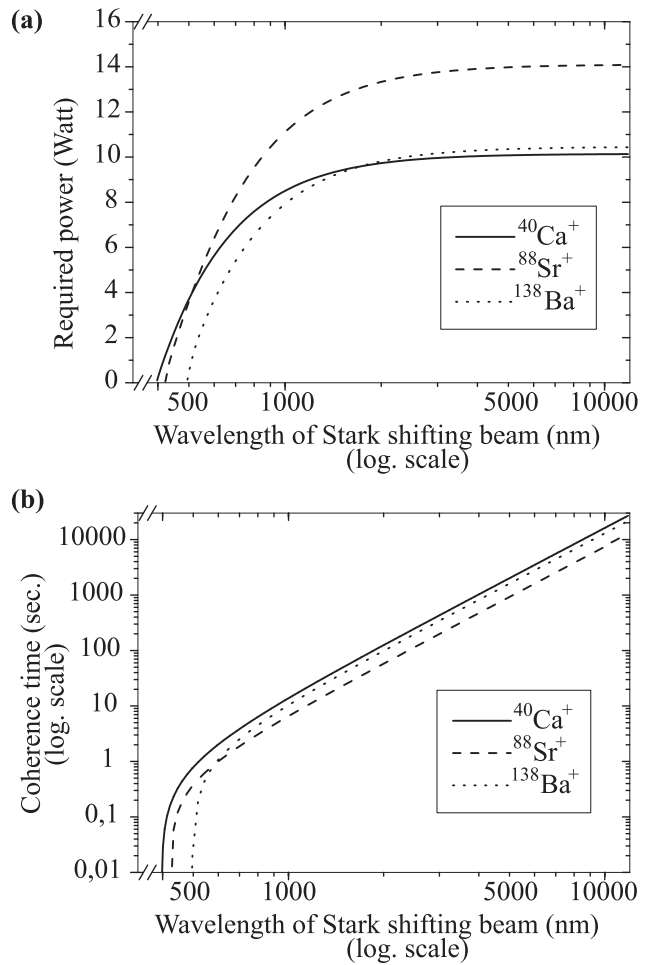}
  \caption{(a) The required laser power as a function of wavelength for
  obtaining an ac Stark shift difference $E_{1,2}=\hbar\omz/2$, when
  $\omz=2\pi\times 1.0\unit{MHz}$ and $W=30\unit{\mu m}$ for \cafp, \srp, and \bap.
  (b) The corresponding coherence time ($\Gamma_{sc}^{-1}$).}
  \label{fig:graphs}
\end{figure}

In Fig. 3a, the laser power required to achieve an ac Stark shift
difference $E_{1,2}=\hbar\omz/2$ in the case of $\omz=2\pi\times
1\unit{MHz}$ is presented for \cafp, \srp, and \bap\ as a function
of the laser wavelength~\footnote{In Fig.~\ref{fig:graphs} we do
take the fine-structure splitting and the different couplings to
the fine-structure levels into account. For the $(n-1)D-n'F$
transitions we sum over $n'=4-10$. The data used are from
Ref.~\citep{NIST-database} (\cafp) and Refs.~\citep{C.E.Moore2,
C.E.Moore3, Lindgaard} (\srp\ and \bap). For a good overview of
the S-, P- and D-levels, see Ref.~\citep{James}.}. The waist of
the laser beam is taken to be $30\unit{\mu m}$, which is much
larger than the equilibrium spacing of $5.6\unit{\mu m}$ ,
$4.3\unit{\mu m}$ and $3.7\unit{\mu m}$ for the \cafp-, \srp-, and
\bap-ions, respectively. The required power, which approaches a
constant in the long-wavelength limit, is well within reach of
commercial lasers, e.g., a $\rm CO_2$ laser
($\lambda=10.6\unit{\mu m}$), a Nd:YAG laser
($\lambda=1064\unit{nm}$), or a frequency-doubled Nd:YAG laser
($\lambda=532\unit{nm}$).

Another very important parameter to consider in the present
scheme, is the spontaneous scattering rate, $\Gamma_{sc}$, of
light from the Stark-shifting laser beam, since it will limit the
ultimate coherence time. Under the assumptions made above in
calculating the ac Stark shifts, the sum of the scattering rates
of both ions can be expressed as:
\begin{eqnarray}\label{eq:sc-rate}
\Gamma_{sc}&&
=\frac{E_{1,2}}{\kappa\psi}\times\frac{e^{-1/2}3\pi c^{2}\omega^{3}}
{\hbar}\nonumber\\
&&\times\Biggl[\frac{1}{\omega_{P}^{6}}\left(\frac{\Gamma_{P}}
{\omega_{P}-\omega}+\frac{\Gamma_{P}}{\omega_{P}+\omega}\right)^{2}\nonumber\\
&&+\sum_{n'}\frac{1}{\omega_{n'F}^{6}}\left(\frac{\Gamma_{n'F}}
{\omega_{n'F}-\omega}+\frac{\Gamma_{n'F}}{\omega_{n'F}+\omega}\right)^{2}
\Biggr],
\end{eqnarray}
where $\Delta z/W\ll 1$, as obeyed by the parameters used in Fig.
3, is assumed~\citep{Grimm-opt}.

In Fig. 3b, the coherence time (or rather $\Gamma_{sc}^{-1}$) is
plotted as a function of laser wavelength, and we see that in the
long-wavelength limit, the coherence time grows as the wavelength
to the third power, which is also readily deduced from
Eq.~\eqref{eq:sc-rate}. Hence, at first, a $\rm CO_2$-laser seems
to be favorable. However, since the lifetime of the
\state{2}{D}{5/2}-level is only $1.0\unit{s}$, $345\unit{ms}$, and
$47\unit{s}$ for \cafp,\srp, and \bap, respectively, the use of
the fundamental wavelength of a Nd:YAG laser might be more
attractive, since this will be much easier to focus to the
required spot size. Actually, in current experimental setups the
maximal coherence time is limited by heating of the ions on a
timescale of $1-100\unit{ms}$~\citep{Turchette-heating, Roos},
hence even a continously operated frequency-doubled Nd:YAG laser
can be used without introducing significant additional
decoherence.

There are several reasons for not choosing $W$ too large compared
with $\Delta z$. First, the required power to achieve a certain
energy difference $E_{1,2}$ grows as $W^{3}$. Second, the total
scattering rate for a fixed $E_{1,2}$ also increases with $W$
($\Gamma_{sc}\propto W$). Furthermore, it should be noted that
although a large $E_{1,2}$ implies a short coherence time, it also
allows a high gate-speed.

The effect of the (internal state dependent) gradient-force
exerted on the ions by the Stark-shifting laser beam has to be
considered. The maximal gradient-force will be on the order of
$F_{grad}=-\partial\varepsilon_{\downarrow}/\partial z\approx
E_{1,2}/\Delta z$. Taking the example of \cafp, and using the same
parameters as above, the maximal gradient-force will be $\sim
10^{5}$ times smaller than the confining force exerted by the
trap, and the associated change in the equlibrium distance
bewtween the ions, $\delta z$, is $\sim 300$ times smaller than
the spread of the vibrational wavefunction. This displacement is
totally negligible. Nevertheless, when the Stark-shifting laser
beam is turned on, an ion obtains a speed $v\approx\delta
z/t_{rise}$, where $t_{rise}$ is the ``rise-time'' of the
Stark-shifting laser beam. The associated kinetic energy must be
much smaller than $\hbar\omz$, which is fulfilled if $t_{rise}\gg
1\unit{ns}$. In practice, this is no limitation.

Above we considered in detail the simple case of two ions and one
motional mode. If we take both motional modes, i.e., the so-called
center-of-mass mode at frequency \omz\ and the stretch mode at
frequency $\sqrt{3}\,\omz$~\citep{James}, into account, the
optimal value of $E_{1,2}$ is slightly changed, but our
conclusions remain valid. Further, we can generalize Case A and
Case B of Fig.~\ref{fig:idea}b to more than two ions. In Case A,
the additional energy-levels will make it difficult to address
individual ions, but it should be possible with a few ions,
particularly in a relatively tightly confining trap. Case B works
just as well with more than two ions. Only, $E_{1,2}/\hbar$ should
not coincide with the frequency of one of the higher motional
modes.

It should be noted that the different ionic
transition-frequencies, while applying the Stark-shifting laser
beam, leads to a differential phase-development of the various
ions. Since the frequency-differences are known, this can be
accounted for by controlling the phase of the addressing
light-field.

\begin{figure}
  \centering
  \includegraphics[width=\linewidth]{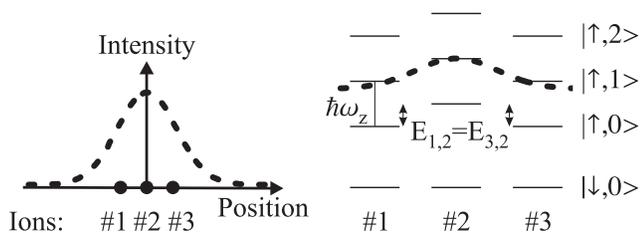}
  \caption{A Stark-shifting laser beam making two ions
  have the same unique resonance-frequency. This allows for selective
  addressing of any pair of ions for two-qubit operations.}\label{fig:MS}
\end{figure}

In addition to individual addressing, a Stark-shifting laser beam
can be used for realizing two-ion quantum logic operations with a
single bichromatic laser pulse, as proposed by M\o lmer and S\o
rensen~\citep{Klaus-entanglement, Anders}, between \emph{any} two
ions in a string. As an example, we show in Fig.~\ref{fig:MS}, how
one can make two ions in a three-ion string have the same unique
resonance-frequency, needed for making a M\o lmer-S\o rensen gate
between these two ions.

The position-dependent ac Stark shift method discussed above is
also applicable to other qubit-levels and ions. For example the
two Zeeman-sublevels of the ground state in \cafp, \srp, or \bap\
can be used as qubit-levels
($\down=\state{2}{S}{1/2}(m_{J}=-1/2)$,
$\up=\state{2}{S}{1/2}(m_{J}=+1/2)$) with qubit-operations
performed by two-photon stimulated Raman transitions. An ac Stark
shift can be induced by a circularly polarized Stark-shifting
laser beam with wavelength $\lambda$ tuned in between the two
fine-structure levels of the excited $P$-state. If we take
$\omz=2\pi\times 1\unit{MHz}$ and $W=30\unit{\mu m}$, as in the
previous discussion, an ac Stark shift difference
$E_{1,2}=\hbar\omz/2$ can be obtained for \cafp\ with a minimum
scattering rate of $161\unit{Hz}$ using a laser with
$\lambda=395.2\unit{nm}$ and a power of $64\unit{mW}$. This
scattering rate allows only for a very limited number of
gate-operations, even in the case where the Stark-shifting laser
beam is only present during the quantum logic processing. For
\srp\ and \bap\ somewhat lower scattering rates can be obtained,
owing to their larger fine-structure splitting. Applying the same
approach to \mgp\ or \bep\ (with hyperfine-levels of the ground
state as qubit-levels~\citep{Sackett-entangled}) is impracticable,
due to their relatively small fine-structure splitting.

In conclusion, we have shown that individual or selective
addressing of trapped ions can be achieved, without introducing
significant decoherence, by utilizing a Stark-shifting laser beam
with modest focusing- and power-requirements. The presented scheme
readily makes it possible to perform single- as well as
multi-qubit gates.

The authors are grateful to Torkild Andersen, Klaus M\o lmer, and
Christophe Salomon for fruitful discussions. This work was
supported by the Danish National Research Foundation through
QUANTOP - the Danish Quantum Optics Center, and the Carlsberg
Foundation.

\bibliography{staanum}
\end{document}